\RequirePackage{fix-cm}
\documentclass[smallextended,natbib]{svjour3}       
\smartqed  
\usepackage{graphicx}
\usepackage{amssymb,amsmath}      
\journalname{Climate Dynamics}
\begin{document}

\title{Chaotic signature of climate extremes\thanks{Ogunjo S. T. acknowledges funding and support from the Max Planck Institute for the Physics of Complex Systems}
}


\author{Ibiyinka FUWAPE         \and
        Sunday OLUYAMO          \and
        Babatunde RABIU         \and
        Samuel  OGUNJO 
}


\institute{I. Fuwape, S. Oluyamo \at
              Department of Physics, Federal University of Technology, Akure \\
              Ondo State, Nigeria.\\
              \email{iafuwape@futa.edu.ng, ssoluyamo@futa.edu.ng} \\
             \emph{Present address:} of I. Fuwape  \\
             Michael and Cecilia Ibru University, Ughelli, Delta State, Nigeria.
           \and
           B. Rabiu \at
              Centre for Atmospheric Research \\
              National Space Research and Development Agency,\\
              Anyigba, Kogi State, Nigeria.
           \and
           S. T. Ogunjo \at
           Department of Physics, Federal University of Technology, Akure \\
           Ondo State, Nigeria. \\
           \email{stogunjo@futa.edu.ng, ogunjo@pks.mpg.de}.}
\date{Received: date / Accepted: date}

\maketitle

\begin{abstract}
Understanding the dynamics of climate extreme is important in its prediction and modeling.  In this study, linear trends in percentile, threshold, absolute, and duration based  temperature and precipitation extremes indicator were obtained for the period 1979 - 2012 using the ETCCDI data set.  The pattern of trend was compared with nonlinear measures (Entropy, Hurst Exponent, Recurrence Quantification Analysis) of  temperature and precipitation.  Regions which show positive trends in temperature based extremes were found to be areas with low entropy and chaotic.  Complexity measures also revealed that the dynamics of the southern hemisphere differs from that of the northern hemisphere.

\end{abstract}

\section{Introduction}\label{intro}
The impact of climate change will be felt in varying degrees in different parts of the world.    Global mean surface temperature was reported to rise about $0.6^o \,C$ in the last century and expected to rise between $1.4 - 5.8^o\,C$ in the next century \citep{boo2006}.  The observed trends in climate time series can either be deterministic (external forcing) or stochastic (intrinsic variability)  \citep{franzke}.  At the regional level where intrinsic climate variability play a great role, climate extremes in temperature and precipitation have significant impact in our agriculture, health, energy, ecosystem and society \citep{schoof2016}.  A set of indices, for monitoring of climate extremes based on temperature and precipitation data, has been defined by the Expert Team on Climate Change Detection and Indices (ETCCDI).   The ease of use and interpretation of the defined indices have made it useful for the study of global extreme events \citep{sillman2013,alexander2006} and for regions \citep{schoof2016,alexander2017,boo2006,tank2003}.  Climate change is expected to modify the characteristics of extreme weather and climate events \citep{abiodun2017}.  For instance, increasing warm and decreasing cool temperature based indices have been reported for Australia \citep{collins} and Southeast Asia \citep{manton2001}, increase of about $3\%$ per decade in short duration extreme precipitation events in the United States  \citep{kunkel1999} and increased occurrence of extreme temperature over Nigeria \citep{abiodun2013,abatan2017}.

According to \citet{Sivakumar2004}, it is important to investigate complexity in geophysical phenomena for short and long term predictions, model testing and better description of the phenomena.  The development of nonlinear techniques and their robust performance over linear methods have led to an increase in the search for structures and understanding of complexity in geophysical phenomena in recent times \cite{sivakumar2007}. Detecting chaos is more complicated in time series than in flows and discrete dynamical systems.  Graphical methods of identifying chaotic systems includes phase space reconstruction, poincare section, close return plot, and recurrence plot.  New measures of complexity have been introduced based on structures in a recurrence plot \citep{Zbilut2015}.  Complexity and level of chaos can be estimated by computing the dimension and entropy of the system \citep{rosenstein1993}.  The sensitivity of a system to initial conditions can be determined by the Lyapunov exponents.  A system with at least one positive Lyapunov exponent is regarded as being chaotic.  Different algorithms for computing the Lyapunov exponent has been proposed by \citet{rosenstein1993}, \citet{sano1985}, \citet{kantz1994} and \citet{wolf1985}.  Other methods of identifying chaotic systems include the method of surrogate data \citep{theiler1992}, $0-1$ test \citep{Gottwald2004} and delay vector variance \citep{Fuwape2016,Gautama2004}.   Chaos investigation has been conducted into precipitation data \citep{Fuwape,Kyoung2011,Sivakumar2006}; temperature data \citep{Fuwape,Millan2010},  wind speed data \citep{an2012,Samet2014} and solar radiation data \citep{Ogunjo2017}.  The presence of noise, large number of zeros, and data size has been cited as limitations to the use of nonlinear techniques.  The presence of zeros was deemed to be less severe when compared to presence of noise and data size in the application of nonlinear techniques to other fields \citep{Sivakumar2006}.  In the study of climate over a period of thirty years, the data length is sufficient for extraction of features \citep{Sivakumar2005}.



In the case of El Ni$\tilde{n}$o Southern Oscillation (ENSO), the discussion is about whether a deterministic or stochastic model best describes it.    Rejection of chaos in the time series of ENSO  has been disputed based on analysis carried out using Largest Lyapunov Exponent \citep{kawamura}, Correlation Dimension \citep{kawamura}, close return plots \citep{ahn2005} and determinism \citep{binder2002}.   The theory  that ENSO time series is a stochastic system rather than a chaotic one has led to the development of stochastic approach for ENSO \citep{ubilava,hall2001}.   Several studies have identify chaos in the dynamics of ENSO using tools such as bred vector \citep{tang2010}, dynamical models \citep{vallis1986,chang1996,tziperman},  false nearest neighbour and correlation dimension \citep{chang1996,tsonis2009}, Lyapunov Exponent \citep{tsonis2009} and nonlinear prediction error \citep{elsner1993}.   The dimension of ENSO reported from dynamical models include 3.5 \citep{tziperman}, 2.4 \& 2.6 \citep{chang1996}, $<2$ \citep{jin2005}, 3 and 4 \citep{tsonis2009} and 8 \citep{aijain1998}.   The possibility of co-existing chaotic dynamics in the seasonal dynamics and stochastic dynamics in the interannual timing and strength of ENSO  \citep{zivkovic} and noise induce chaotic dynamics \citep{stone1998} has been proposed.

It has been posited that our climate system is both chaotic and complex \citep{rind}.  For example, ENSO has been shown to influence river flow in the tropics and subtropics  \citep{khan2006} and Asian irrigation has remote impact on African rainfall \citep{vrese}.   The effect of changing climate in one region on or in relation to another region can only be studied on a global rather than regional level.  While several studies have indicated chaos in geophysical phenomena at local level, the dynamics of our changing climate needs to be investigated on the global scale.  To this end, this study aims to investigate the chaotic signature of climate change by comparing dynamical analysis of temperature and precipitation to observed trend across the world.

\section{Data}\label{sec:data}

\subsection{Data}\label{sec:2}
The indices described by the ETCCDI \citep{tank2009} can be categorized into percentile based indices, absolute indices, threshold indices, duration indices and other indices \citep{alexander2006}.   An index was chosen in each category for both temperature and precipitation as shown in Table \ref{tab1}.  Annual values of all ETCCDI indices have been computed for various data sets (global climate models, observational data and reanalyses) using consistent methodologies \citep{sillman2013}.  Computed values of indices for the ERA-Interim reanalysis data set from 1979 - 2012 were retrieved from the ETCCDI indices archive at {www.cccma.ec.gc.ca/data/climdex/climdex.shtml}.  Linear trends in the climatic extremes over 34 years were taken as the linear regression coefficient.
\begin{table*}
  \centering
  \caption{List of extreme precipitation and temperature indices as defined by the ETCCDI \citep{sillman2013} used in this study}
  \begin{tabular}{llp{6.5cm}}
    \hline\noalign{\smallskip}
    Index  & Label & Definition \\  \hline
    Extremely wet days  & R99p  &  Annual precipitation derived from days $>$ 99th percentile\\
    Max. 1-day precipitation & RX1day & Maximum 1-day precipitation \\
    Heavy precipitation days  & R10mm & Annual count of days with precipitation $>$ 10 mm\\
    Consecutive wet days & CWD & Max. number of consecutive days with precipitation $>$ 1 mm\\
    Simple daily index & SDII & Mean precipitation amount on wet days\\
    Warm days & TX90p  & Percentage of days with $T_{max} >$  historical 90th percentile value\\
    Maximum $T_{min}$ & TNx & Maximum value of $T_{min}$\\
    Summer days  &  SU  &  Number of days with $T_{max} > 25^oC$\\
    Warm spell duration & WSDI  & Annual count of at least six consecutive days with $T_{max} >$ the historical $90th$ percentile value\\
    Diurnal range & DTR&  Mean difference between daily $T_{max}$ and $T_{min}$\\
    \hline
  \end{tabular}
  \label{tab1}
\end{table*}

In order to ensure consistency and uniformity, data for computation of nonlinear measures were also obtained for daily precipitation and minimum temperature data from the ERA-Interim reanalysis studies \citep{Dee2011} from 1979 - 2012 on a regular $1.5^o \times 1.5^o$ grid.   However, data processing such as removal of outliers and consecutive equal values control were not carried out on the data.  Approaches to removal of trends have been shown to have effect on computed nonlinear measures \citep{ogunjo2015effect}, hence, precipitation and minimum temperature anomalies were used for nonlinear analysis.
\section{Methodology}\label{sec:method}

\subsection{Entropy}
Entropy is regarded as a measure of disorder in a system.  In thermodynamics, entropy is the degree of disorder in a system, in information theory, it is the amount of information obtainable from a system while in statistical mechanics, entropy is the number of microscopic state of a system in thermodynamic equilibrium.  The concept of entropy has been extended as a measure of complexity in a time series.  The Boltzmann-Gibbs entropy is defined as
\begin{equation}\label{shannon}
    S_l = -\sum_{i=1}^W p_i\ln(p_i)
\end{equation}
where $W$ is the number of macroscopically indistinguishable microscopic configurations \citep{singh1997}.  \citet{Tsallis1988} defined entropy for nonextensive system as
\begin{equation}\label{tsallis}
    S_q = \frac{k}{q-1}\left( 1 - \sum_{i=1}^W p_i^q \right)
\end{equation}
where $p_i$ are the probabilities associated with the microscopic configurations, $W$ is their total number, $q$ is a real number, and $k$ is Boltzmann's constant \citep{Kalimeri2008}. In the limit $q\rightarrow1$, Tsallis entropy (equation \ref{tsallis}) becomes the Boltzmann-Gibbs entropy (equation \ref{shannon}).   Entropy analysis has been used in the spatio-temporal dynamics of tropical climate \citep{Fuwape}, delineation of water resources zones in Japan \citep{kawachi2001}, pre-seismic emissions \citep{Kalimeri2008}, investigation of abrupt climate change in Earth's climate \citep{Gonzalez2011} and application in hydrological processes \citep{singh1997}.

\subsection{Hurst Exponent}

Given a time series $Z$, the average $X(\tau)$ and standard deviation $S(\tau)$ of the sequence of observations are computed as Equation \ref{rs1} and \ref{rs2} respectively (where $\tau$ is the time lag).

\begin{align}
    X(\tau) &= \sum_{u=1}^t (Z(u) - \langle Z\rangle_\tau) \label{rs1}\\
    S(\tau) &= \left(\frac{1}{\tau} \sum_{t=1}^\tau (Z(t) - \langle Z\rangle_\tau)^2 \right)^{\frac{1}{2}} \label{rs2}
\end{align}
Defining the self adjusted range as
\begin{equation}\label{rs3}
    R(\tau) = \max_{t=1}^\tau \{ X(\tau) \} - \min_{t=1}^\tau \{ X(\tau) \}
\end{equation}
Hurst exponent, H is then computed from the expression $\log(R/S)=\log(\tau/2)^H$ \citep{Fuwape2016}.  Persistence, anti-persistence, and uncorrelated processes are indicated by $0.5<H<1$, $0<H<0.5$ and $H=0.5$ respectively.   Hurst exponent has been applied to precipitation and river runoff records using DFA \citep{Kantelhardt2006}, annual precipitation in the United States using Rescale Range method \citep{potter1979} and other time series \citep{Fuwape2016}.
\subsection{Recurrence Quantification Analysis (RQA)}
\citet{marwan2007} defined the recurrence of a system $R_{i,j}$ at time $i$ in a different time $j$ as
\begin{equation}\label{rqa1}
    R_{i,j}^{m,\epsilon_i} = \theta(\epsilon_i - ||\vec{x}_i-\vec{x}_j||)
\end{equation}
where $\epsilon_i$, $||\cdot||$ ,$\theta$ are the threshold for distance, Euclidean norm and Heaviside function respectively \citep{Panagoulia2014}.  Recurrence plots have small scale structures which can be used to quantify complexity.  Three of these measures, known as recurrence quantification analysis will be employed in this study:  Recurrence rate (REC), Determinism (DET) and Divergence (DIV).  REC is defined as the density of recurrence points, DET is the ratio of recurrence points that form diagonal structures to all recurrence points and DIV is the inverse of the longest diagonal line in the recurrence plot.  Relationship between RQA parameters and measures such as Lyapunov exponent, correlation dimension and Renyi entropy has been reported in literature \citep{marwan2007,Zbilut2006}.  In this study, the threshold is taken as $\epsilon > 5\sigma$.  Recurrence plots and recurrence quantification analysis have been employed in several climate studies \citep{Panagoulia2014}.

\section{Results and discussion}
In this section, the results of linear trends in temperature and precipitation trends, as well as chaotic analysis carried out as described in Section \ref{sec:method}, on minimum temperature, maximum temperature and precipitation data are presented and discussed.

\subsection{Linear trend in extreme temperature and precipitation indices}
\begin{figure}
\centering
  \includegraphics[scale=0.28]{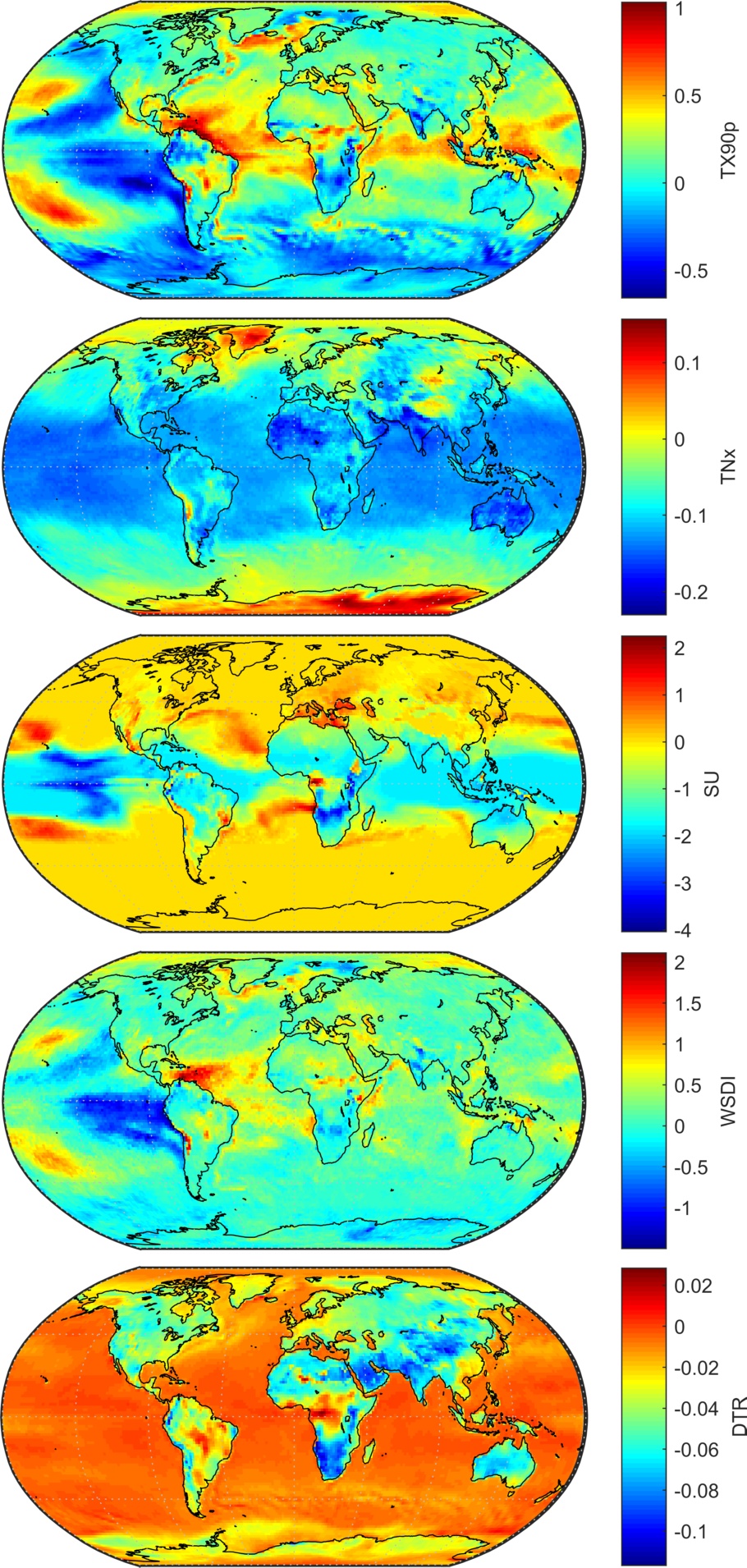}\\
  \caption{Trends in temperature indices  (from top to bottom) TX90p, TNx, SU, WSDI and DTR}\label{fig1}
\end{figure}

Figure \ref{fig1} shows the trend in temperature based extremes over the globe in the period 1979 - 2012.   Over terrestrial land surface, the percentile based index, TX90p showed a decreasing trend in the northern coast of South America, southern America, southern Asia, western Australia and Canada.  Continental warming could be seen in the central regions of South America and Africa.  Over the Ocean,  the El Nino region in the Pacific showed decreasing trend while warming could be observed in the Atlantic and Indian Oceans.   The absolute index, TNx showed decreasing trends with varying intensity across the tropics, central Europe and continental North America.  It also showed warming in Antarctica and the Arctic region.  The number of summer days (SU) showed warming in the Mediterranean sea and Europe.  Across the tropics, cooling could be observed in both land and sea surface with regions of warming in the northern and southern regions of both Pacific and Atlantic oceans.  Warm spell duration (WSDI) showed similar trend to the percentile index (TX90p), albeit with a lesser degree of intensity.  The world oceans, central regions of South America and Africa showed increasing trend in diurnal temperature range.  The greatest decrease in DTR were found in Southern Africa and Asia.  The results obtained show significant similarity with other global \citep{christidis2016,hansen2010,vose2005} and regional \citep{chou2014evaluation}  reports.

\begin{figure}
\centering
  \includegraphics[scale=0.3]{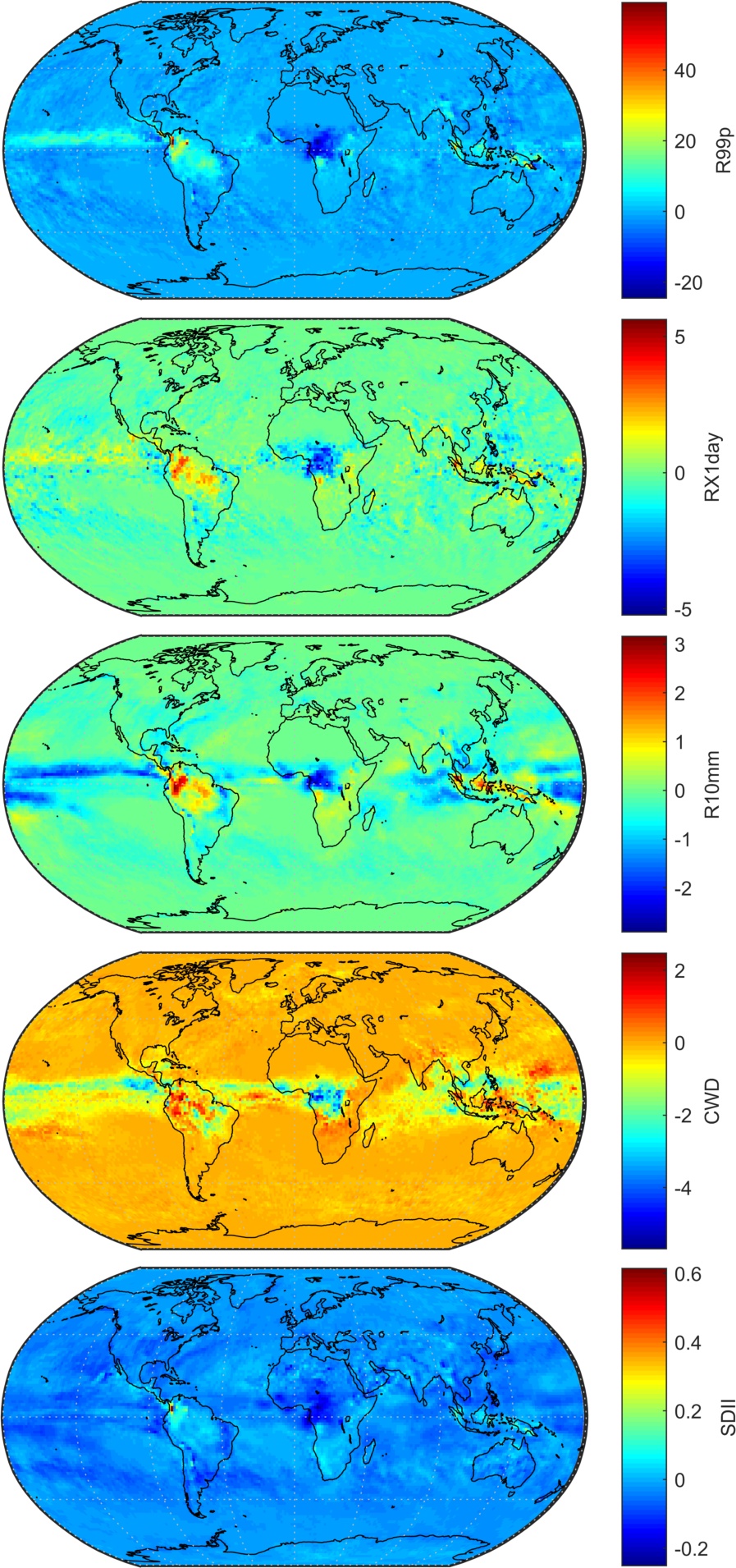}\\
  \caption{Trends in precipitation indices  (from top to bottom) R90p, RX1day, R10mm, CWD and SDII}\label{fig2}
\end{figure}

The percentile (R99p) and absolute (RX1day) indices showed similar trend (Figure \ref{fig2}).  Notable trends were found in central Africa, northern region of South America and pacific islands.  While a decreasing linear trend was observed in central Africa, an increasing trend was found in northern region of South America.  The threshold based index (R10mm) and duration based index (CWD) revealed changes in tropical region.  Along the tropics, a cooling trend was observed in the R10mm index and a warming trend in the northern regions of South America and Pacific Islands.  Similar but more noticeable pattern was observed in the CWD index.

\subsection{Complexity in global temperature and precipitation}
\begin{figure}
\centering
  \includegraphics[scale=0.3]{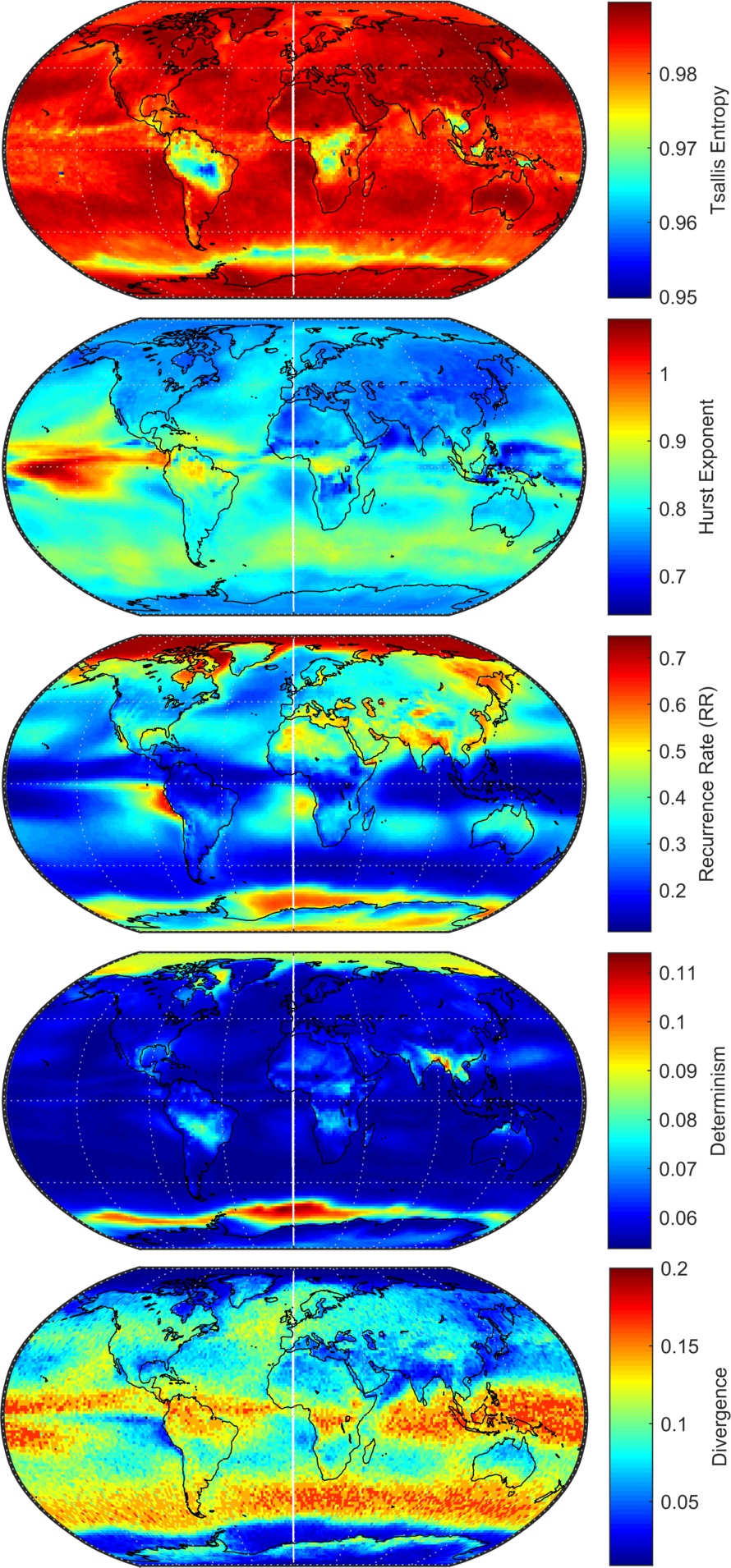}\\
  \caption{Complexity measures for minimum temperature using (from top to bottom) Tsallis entropy, Hurst Exponent, Recurrence Rate, Determinism, Divergence}\label{fig3a}
\end{figure}

\begin{figure}
\centering
  \includegraphics[scale=0.3]{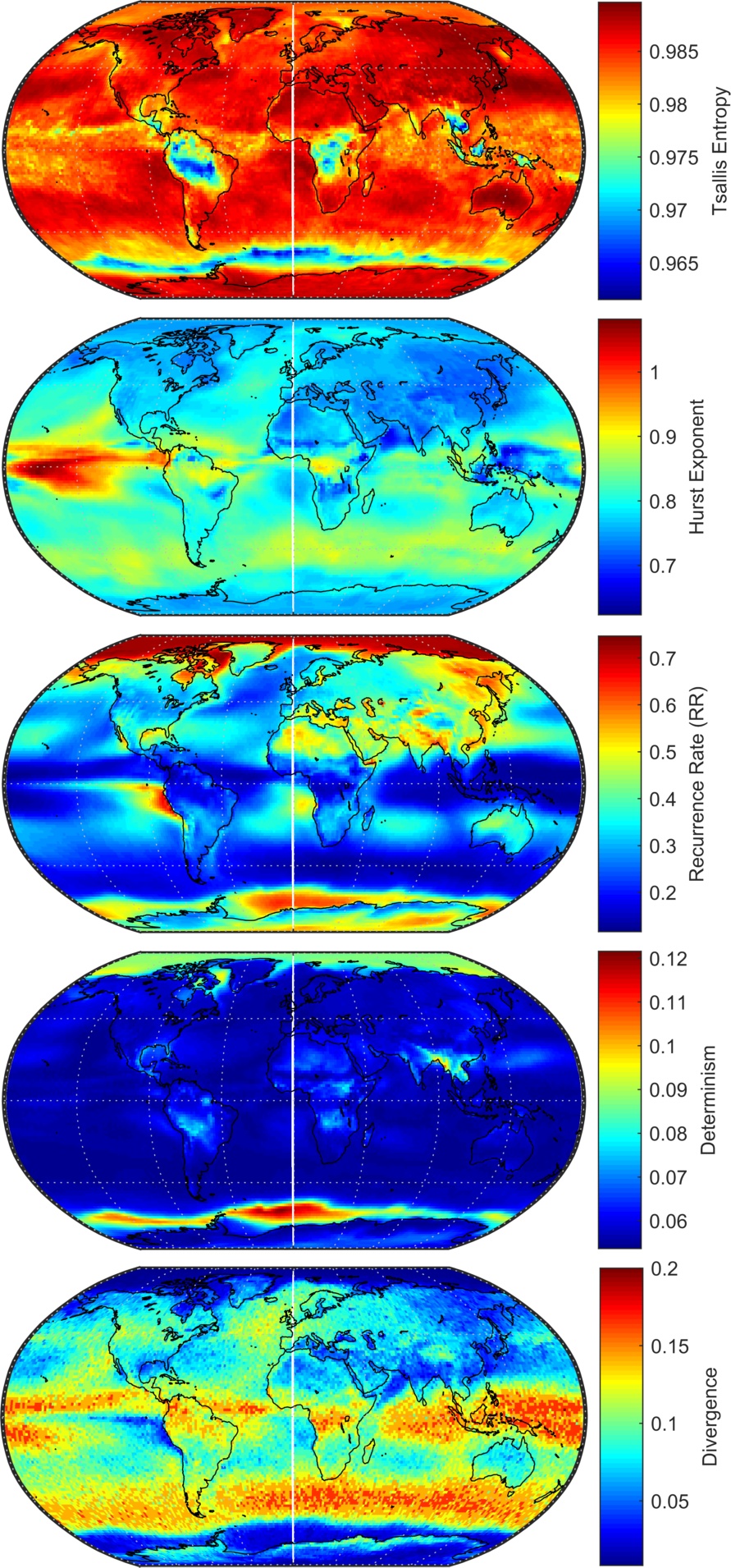}\\
  \caption{Same as in Figure \ref{fig3a} but for maximum temperature}\label{fig3b}
\end{figure}

Figures \ref{fig3a} and \ref{fig3b} showed the computed nonlinear signatures of minimum and maximum temperature respectively using the method of Tsallis Entropy, Hurst Exponent and Recurrence Quantification Analysis (RQA).   The Tsallis Entropy of minimum temperature revealed a high degree of disorderliness or randomness over most of the land.  The entropy values are much higher for northern and southern Africa, most of Asia, Australia and northern region of North America than Europe and southern region of North America.   Lower entropy values were observed for the central region of both South America, Africa and tropical Asia.  Surfaces of the ocean reveal a different dynamics as the tropical regions of the oceans show a lower entropy value than surrounding ocean body.  This is more obvious in the Tsallis entropy value of maximum temperature (Figure \ref{fig3b}).  Both land and ocean in the tropics show lower entropy values than the surrounding areas.  The increasing entropy with increasing latitude seen over West Africa has also been reported by \citet{Fuwape}.

The computed Hurst Exponent for both minimum and maximum temperatures show identical features.  Most of the global land mass show persistence (i.e. an increasing (decreasing) trend will be followed by an increasing (decreasing) trend).  These values are, however, lower than that obtained over the ocean.  The Hurst exponent values in the El nino region shows $1/f$ noise while regions in the Atlantic ocean show low values of Hurst Exponent.  Regions in the central part of South America and Africa showed values different from global trends.

Low values of recurrence rates shows that minimum temperature over the oceans are chaotic except for the western coast of both South America and Atlantic Ocean.  The highest value of recurrence rate were observed in the Arctic Ocean.  Australia shows a chaotic values in minimum temperature which reduces towards the equator.  Land mass along the tropics including the islands in the Pacific Oceans shows chaotic trend while continental land mass tend towards stochastic values.  From the map of divergence, the chaotic nature of minimum and maximum temperature could be ascertained.  Low dimensional chaos could be found in minimum and maximum temperature across continental land mass.  This result could explain the low predictability reported in the El nino region and Pacific islands using observational data \citep{Hunt2017}.

In the Pacific ocean, predominant low entropy values were interspersed  by narrow streams of high entropy values below and above the El nino regions.  The low entropy values observed in the Pacific ocean does not translate into low entropy values in continental North and South America, although streaks of high entropy values transverse longitudinally through the central North America.   Contrary to the low entropy values obtained for minimum and entropy values in central regions of South America and Africa, high entropy values were found for precipitation data in those regions. Similarly, while high entropy values obtained for minimum and maximum temperature in northern Africa and Australia, the regions showed low entropy values for precipitation.  The low entropy values in the Australian continent belongs to the stream of high entropy values observed in the Tropic of Capricorn.

Precipitation showed persistence across the tropics.  The spread seems to weaken as it moves from the Pacific ocean, through central South America into the Atlantic Ocean.  Incoherent patterns of uncorrelated precipitation could be observed across the globe.  In the northern hemisphere, precipitation showed random walk in northern Africa, parts of Europe, Asia and North America.

The dynamics of precipitation across the globe was investigated using recurrence rate, determinism and divergence.   The same trend of results were observed across the three quantifiers with recurrence rate showing more distinct features than determinism and divergence.  Most of the continental land mass showed chaotic behaviour with low recurrence rates with the exception of regions in northern Africa, Australia, Yemen and Oman.  The \textbf{stochastic} nature of precipitation in the Arabian sea did not follow the chaotic trend observed in the Indian ocean.  The region from the Indian ocean through the Pacific islands to the Indian ocean were also found to exhibit chaotic precipitation.  In the El nino region, the chaotic region gave way to two pitchfork surrounding a \textbf{non-chaotic} region.  \textbf{Non-chaotic} values of precipitation were found along the Tropic of Capricorn with  more intensity in the Pacific and Indian Oceans.  In the Atlantic ocean, the coast of West Africa and northern Africa also exhibit \textbf{non-chaotic} behaviours.

\subsection{Relationship between trends of extreme climate and complexity measures}

The footprint of precipitation extremes could not be observed in the global isoentropy or chaos map but the signature of temperature based extremes could be found around the globe. This suggests that solar activity is the main large-scale driver of deterministic chaos across the globe \citep{faranda}.  This position is further strengthened by the different patterns of results in the northern and southern hemisphere.  The differences observed between the northern and southern hemisphere have been attributed to difference in absorbed solar radiation, difference in outgoing long wave radiation and cross-equatorial heat transport \citep{feulner2013}.    The net surface heat flux has been found to account for the sea surface temperature variability over the southern hemisphere \citep{reason2000}.

Considering trends in temperature climate extremes (Figures \ref{fig1})  and corresponding complexity measures in minimum and maximum temperature (Figure \ref{fig3a} and \ref{fig3b}), similarities could be observed between linear trends in DTR and TX90p, and Tsallis entropy for land areas.  Regions with positive trends in DTR corresponds to regions with lower entropy.  These regions include central regions of both South America and Africa and eastern coast of Asia.  Across the tropics, the region off the coast of South America with negative trend in TX90p show corresponding chaoticity (low recurrence rate) while positive trend in the tropical Atlantic Ocean  and Equatorial region show high recurrence rate.  However, the cooling observed in TX90p in the Southern mid-latitude did not result in low recurrence rate.  This can be attributed to the role of ocean dynamics \citep{Gille1275,reason2000}.

\section{Summary and concluding remarks}

We have studied the global spatial trend in temperature extremes (TX90p, TNx, SU, WSDI and DTR) and precipitation extremes (R90p, RX1day, R10mm, CWD and SDII).  The results obtained showed varying degrees of cooling and warming across the earth with enhanced activity in the ENSO and tropical region.  Similarly, the dynamical complexities were investigated using Tsallis Entropy, Hurst Exponent, and Recurrence Quantification Analysis (RQA).  While the footprint of precipitation extremes could not be observed in the analysis, positive trend in temperature extremes were found in regions which show low entropy and chaos. Furthermore, the nonlinear analysis of temperature also revealed the different dynamics of both northern and southern hemisphere, which confirms previous studies.  While we have studied global spatial dynamics of temperature and precipitation, there is the need for temporal studies to understand the evolution of climate phenomena.

\begin{figure}
\centering
  \includegraphics[scale=0.3]{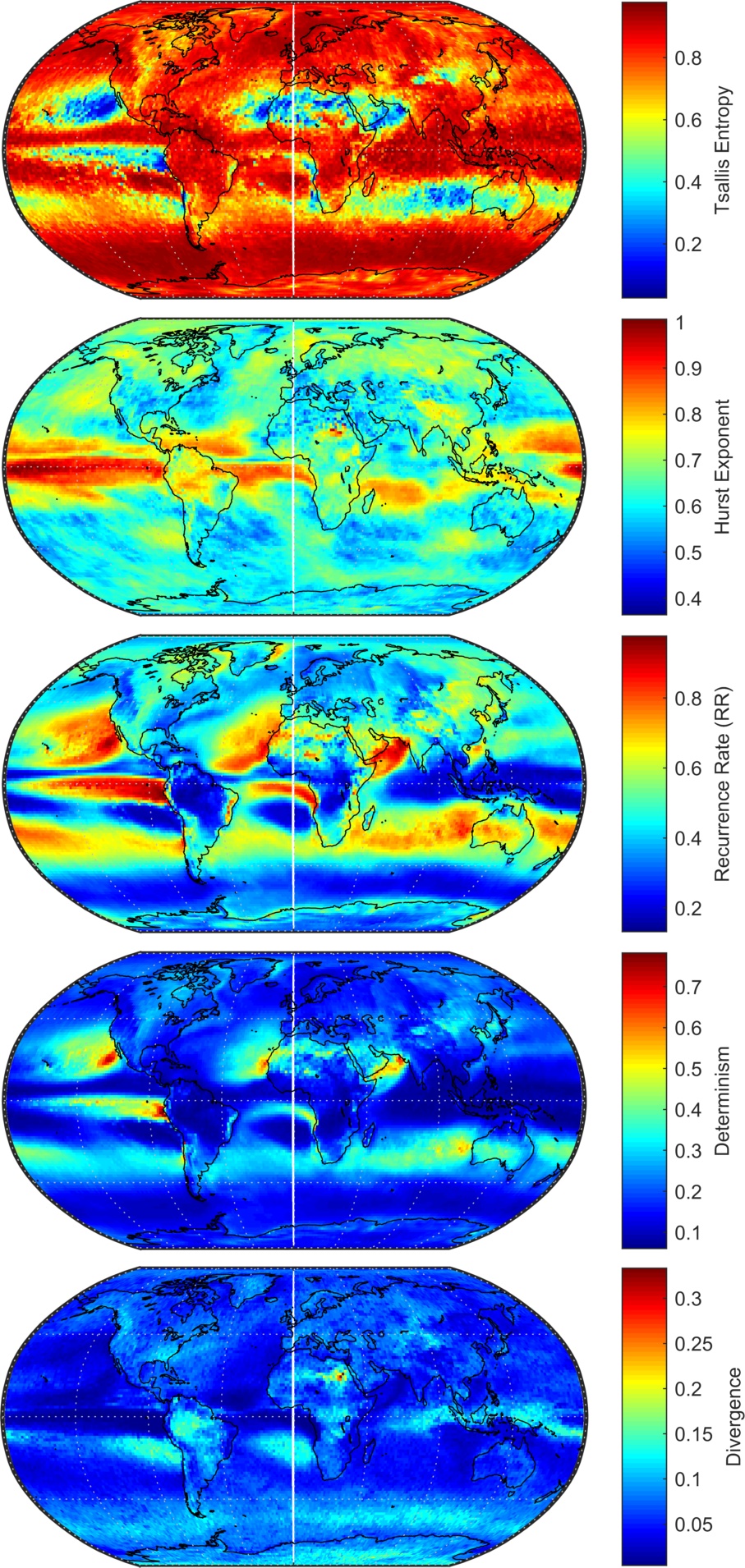}\\
  \caption{Same as in Figure \ref{fig3a} but for precipitation}\label{fig4}
\end{figure}

\begin{acknowledgements}
Part of this research was carried out at the Max Planck Institute for the Physics of Complex Systems, Dresden, Germany by Ogunjo Samuel.
\end{acknowledgements}

\end{document}